\documentclass[%
 reprint,
 superscriptaddress,
 showpacs,preprintnumbers,
 amsmath,amssymb,
 aps,
 prb,
]{revtex4-1}

\usepackage{graphicx}
\usepackage{dcolumn}
\usepackage{bm}

\usepackage{upgreek}
\usepackage{amsmath}

\begin{document}


\title{Strong coupling between excitons and magnetic dipole quasi-bound states in the continuum 
	in WS$_2$–TiO$_2$ hybrid metasurfaces}

\author{ Meibao Qin }
\affiliation{Department of physics, Nanchang University, Nanchang 330031, People’s Republic of China}
\author{Shuyuan Xiao}
\email{syxiao@ncu.edu.cn}
\affiliation{Institute for Advanced Study, Nanchang University, Nanchang 330031, People’s Republic of China}
\affiliation{Jiangxi Key Laboratory for Microscale Interdisciplinary Study, Nanchang University, Nanchang 330031, People’s Republic of China}
\author{Wenxing Liu }
\affiliation{Department of physics, Nanchang University, Nanchang 330031, People’s Republic of China}
\author{Mingyu Ouyang }
\affiliation{Department of physics, Nanchang University, Nanchang 330031, People’s Republic of China}
\author{Tianbao Yu }
\email{yutianbao@ncu.edu.cn}
\affiliation{Department of physics, Nanchang University, Nanchang 330031, People’s Republic of China}
\author{ Tongbiao Wang }
\affiliation{Department of physics, Nanchang University, Nanchang 330031, People’s Republic of China}
\author{ Qinghua Liao }
\affiliation{Department of physics, Nanchang University, Nanchang 330031, People’s Republic of China}


\begin{abstract}
Enhancing the light-matter interactions in two-dimensional materials via optical metasurfaces has attracted much attention due to its potential to enable breakthrough in advanced compact photonic and quantum information devices. Here, we theoretically investigate a strong coupling between excitons in monolayer WS$_2$ and quasi-bound states in the continuum (quasi-BIC). In the hybrid structure composed of WS$_2$ coupled with asymmetric titanium dioxide nanobars, a remarkable spectral splitting and typical anticrossing behavior of the Rabi splitting can be observed, and such strong coupling effect can be modulated by shaping the thickness and asymmetry parameter of the proposed metasurfaces. It is found that the balance of line width of the quasi-BIC mode and local electric field enhancement should be considered since both of them affect the strong coupling, which is crucial to the design and optimization of metasurface devices. This work provides a promising way for controlling the light-matter interactions in strong coupling regime and opens the door for the future novel quantum, low-energy, distinctive nanodevices by advanced meta-optical engineering.
\end{abstract}

\maketitle


\section{\label{sec1}Introduction}
Strong coupling of excitons to optical microcavity has received tremendous interest for its fundamental importance in basic quantum electrodynamics at nanoscale and practical applications towards quantum information processing\cite{Liu2014,Kang2018,Zhao2020}. When the coherent exchange rate between the exciton and the optical microcavity is greater than each decay rate, the interaction enters into the strong coupling regime, forming an exciton-polaron and leading to Rabi splitting and anticrossing behavior in the optical spectra\cite{Han2018,Huang2020,Xie2020}. Exciton-polaritons, quasi-particle in a hybrid light-matter state, have attracted a lot of researches activity over the past decade for their promising potential as a designable, low-energy consumption in the application of quantum computing and quantum emitters\cite{Deng2010,Zhang2018,Xie2020a}. The ability to manipulate the strong coupling is elementary to the design of photonic devices. The most basic description of light-matter interaction is given by the coupling intensity $g$. In the dipole approximation, $g = \mu  \cdot E \propto \frac{1}{V}$, $\mu$ represents the transition dipole moment and $E$ represents the local electric field intensity\cite{Toermae2014,Baranov2017,Hugall2018}. In this sense, transition-metal dichalcogenides (TMDCs) have garnered much attention owing to its direct band gaps, large exciton transition dipole moment and exciton response even at room temperature due to quantum confinement in the atomic layer\cite{Wang2018,Debnath2017,Sychev2018}.

Over the past decade, the strong coupling between the excitons of TMDCs and optical microcavities was mostly realized by metallic nanocavities supporting surface plasmon polaritons, which can strongly confine the electric field in ultrasmall mode volume\cite{Bellessa2004,Li2019,Wang2020}. However, the metal has thermal instability in visible region due to large ohmic loss. The Fabry-perot (F-P) cavity constructed by Bragg reflector can realize strong coupling but the integration is difficult and the volume of whispering gallery modes is large, both of which are difficult to be applied in reality\cite{Wei2015,Flatten2016,Wang2016,Chen2017,Liu2017}. Recently, the guided resonance coupled with WS$_2$\cite{Cao_2020}, two-dimensional dielectric photonic crystal slab with WS$_2$ have been successfully reported to achieve strong coupling between dielectric and excitons\cite{Chen2020}. However, the traditional microcavity is difficult to further compress the volume due to the limit of diffraction which greatly affects the local electric field intensity. Another prospect dielectric metasurface will further minimize the volume to enhance the strong coupling. As far as we know that there are few studies on the strong coupling between TMDCs excitons and the resonance in emerging optical metasurface structures. In fact, metasurfaces can support very high diversity of resonance modes, confine the incident light into deep subwavelength volume, and enhance the light-matter interaction at the nanoscale, thus providing a versatile platform for controlling exciton coupling\cite{Jiang2018,Xiao_2020,Mupparapu2020,Sarma2020}.

 In this paper, for the first time, we investigate the metasurface-enhanced strong coupling between excitons in TMDCs and bound states in the continnum resonance (BIC). In the hybrid structure consisting of WS$_2$ and titanium dioxide (TiO$_2$) nanobars, the magnetic dipole (MD) resonance governed by quasi-BIC is obtained by breaking the C$_2$ symmetry and analyzed using the finite element method (FEM), which provides the ideal number of photons to interaction with exciton. A remarkable spectral splitting of 46.86 meV and typical anticrossing behavior of the Rabi splitting can be observed in the absorption spectrum, which can be well described by coupled-mode theory (CMT). By further changing the asymmetry parameter and varying the thickness of the TiO$_2$ metasurface, it is found that the balance of line width of the quasi-BIC mode and local electric field enhancement should be reached to obtain the large Rabi splitting. Our work set an example for strong coupling in TMDCs/metasurface hybrid system and show great flexibility with diverse geometric configurations and different 2D TMDCs materials, which opens an avenue for smart design of novel integrated quantum devices.
 
\section{\label{sec2}Structure and model}

\begin{figure}[htbp]
	\centering
	\includegraphics
	[scale=0.45]{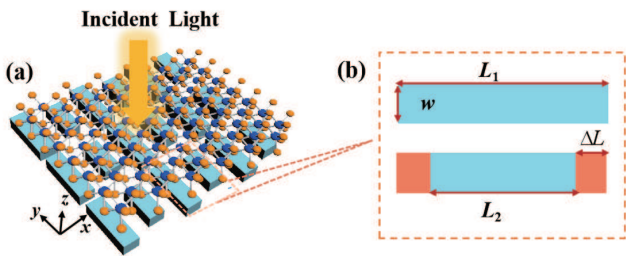}
	\caption{\label{fig1}(a) The sketch of TiO$_2$ metasurfaces with a WS$_2$ monolayer placed on the top. (b) The top-view of unitcell schematic. The TiO$_2$ metasurfaces have multiple design parameters, including the period $P$, length $L_1$ and $L_2$, total thickness $H$, etc. A quasi-BIC resonant wavelength can be adjusted by varying the parameters of $H$ and $L_2$.}
\end{figure}

The proposed hybrid construction, as illustrated in  Fig. \ref{fig1}(a), is composed of a monolayer WS$_2$ lying on the titanium dioxide (TiO$_2$) metasurfaces. In the absence of WS$_2$,  the bare metasurfaces consist of a pair of parallel, geometrically asymmetric nanobars, as depicted in Fig. \ref{fig1}(b), the period of unit cell is $p=450$ nm in both $x$ and $y$ directions, the width of nanobars is $w=100$ nm and a fixed separation between any two neighboring bars is $w_a=125$ nm. The length of the long nanobar is $L_1=400$ nm, while the length of the short nanobar $L_2$ is variable which can generate quasi-BIC mode. The thickness $H$ of the nanobars is also adjustable to match the exciton wavelength. Such metasurfaces can open a radiation channel via introducing an in-plane perturbation in the nanobar length with an asymmetric parameter defined as $\delta  = \Delta L/{L_1}$. Further, we consider a homogenous background with permittivity 1, and choose TiO$_2$ as the constituent material which has high refractive index and negligible absorption loss in the range of visible light. For simplification, the index of TiO$_2$ is assumed as n=2.6. 

\begin{figure}[htbp]
	\centering
	\includegraphics
	[scale=0.35]{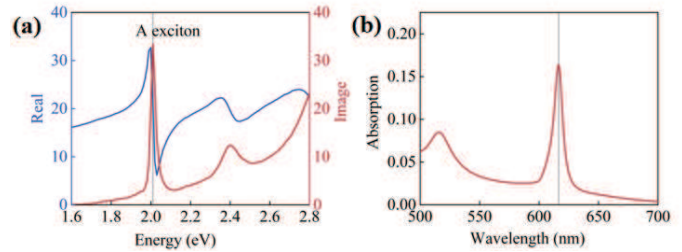}
	\caption{\label{fig2}\ (a) The real (blue line) and imaginary (red line) parts of the permittivity of the monolayer WS$_2$. (b) The absorption curve of the monolayer WS$_2$. A sharp peak value around 616.2 nm (2.014 eV).}
\end{figure}

The permittivity of WS$_2$ is modeled by the Lorentz oscillator model with the thickness of 0.618 nm, adopted from the experimental parameters by Li et al\cite{Li2014},  as shown in Fig. \ref{fig2}(a). The imaginary part has a sharp increase value (red line) around 2.014 eV (616 nm), which is the exciton of WS$_2$ shown in Fig. \ref{fig2}(b), indicating that WS$_2$ has a large line width at 2.014 eV and is suitable for being strong coupling material. In theory, it is likely to reach strong coupling when the resonance wavelength of quasi-BIC draws near the exciton wavelength (616 nm) of the monolayer WS$_2$, and the FEM is used to verify the predication. The thickness of nanobars and the length of short nanobar are initially set with                                                                                                                                                                                                                                                                                                                                                                                                                                                                                                                                                                                                                     $H=85$ nm and $L_2=280$ nm, respectively. In the numerical simulations, the transverse electric (TE) polarized plane wave is normally incident along the $z$ direction, and the periodic boundary conditions are utilized in the $x$ and $y$ direction, and the perfectly matched layers are employed in the $z$ direction.
\section{\label{sec2}Results and discussions}
\subsection{The magnetic dipole quasi-BIC resonance in the TiO$_2$ metasurfaces}	

\begin{figure}[htbp]
	\centering
	\includegraphics
	[scale=0.33]{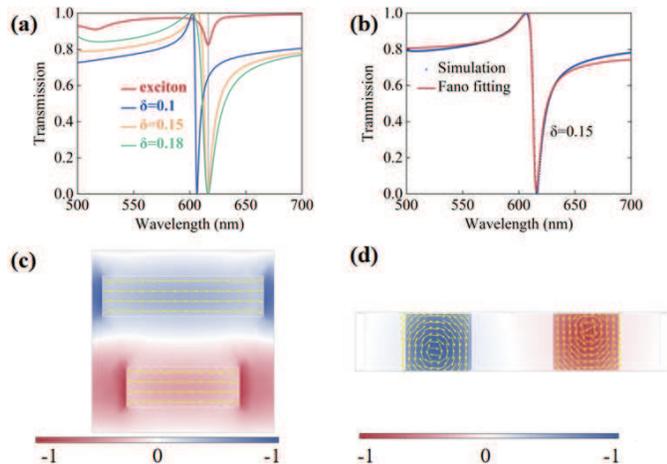}
	\caption{\label{fig3}\ (a) The transmission curves of individual WS$_2$ (red line) and uncoupled quasi-BIC (black line). (b) The red curve and blue curve correspond to theoretical and numerical simulation, respectively. (c) Magnitude of the $y$-component of electric field with asymmetric parameter $\delta {\rm{ = }}0.15$ and $H$=85 nm at the resonance wavelength 616 nm, and arrows indicate the direction of displacement current. (d) Corresponding magnitude of the $y$-component of displacement current and arrows indicate the direction of magnetic field.}
\end{figure}

To obtain a clearer insight into the physics of magnetic dipole quasi-BIC resonance in the metasurfaces, we analysis the transmission spectrum of the TiO$_2$ metasurfaces with different asymmetry parameters are shown in Fig. \ref{fig3}(a), which manifests a Fano lineshape resonance as a result of the in-plane symmetry breaking of the unit cell. In our work, when a perturbation is introduced into an in-plane inverse symmetric $\left( {x,y} \right) \to \left( { - x, - y} \right)$ of a structure, BIC will transform into quasi-BIC and build the radiation channel between a nonradiative bound state and the free space continuum, at the same time, confine part of their electromagnetic field inside the structure is shown in  Figs. \ref{fig3}(b) and \ref{fig3}(c). We take the asymmetric parameter                                                     $\delta {\rm{ = }}0.15$ and $H$=85 nm at the resonance wavelength 616 nm for example, the inverse phase with almost equal amplitude of electric field can be observed in Fig. \ref{fig3}(b), and the circular displacement current in the nanobars generates an out-of-plane magnetic field as shown in Fig. \ref{fig3}(c), which reveal the properties of a magnetic dipole with a strongly localized electrical field inside the nanobars. Then a Fano line will be caught in the transmission spectrum due to the interference between the magnetic dipole and free space continuum. This capture pattern provides a platform for enhancing light-matter interaction at the near-field\cite{Zhang2013,Koshelev2018,Wang2020a}.

We then fit the transmission spectrum $T(w)$ by the Fano formula\cite{Wu2014,Yang2014,Li2019a}
\begin{equation}
{T_{Fano}}\left( w \right){\rm{ = }}{\left| {{a_{\rm{1}}}{\rm{ + }}j{a_{\rm{2}}} + \frac{b}{{w - {w_0} + j\gamma }}} \right|^2},\label{eq1}
\end{equation}
where $a_1$, $a_2$, and $b$ are the constant real numbers,
 ${w_0}$ is the resonant frequency and $\gamma$ is the dissipation of the quasic-BIC, as depicted in the Fig. \ref{fig3}(b).

\subsection{The quasi-BIC resonance and Exciton coupling }

\begin{figure}[htbp]
	\centering
	\includegraphics
	[scale=0.34]{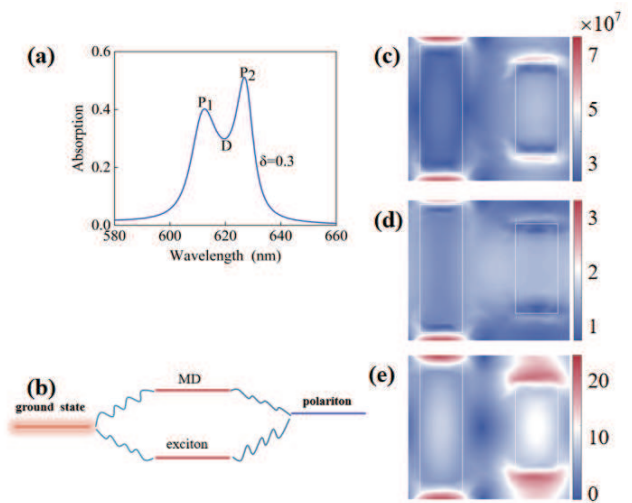}
	\caption{\label{fig4} (a) The absorption curves of the hybrid structure. (b) Two-level coupled oscillator model. (c)-(e) are the electric field distributions of the new hybrid state at absorption peaks labeled by P$_1$,   P$_3$ and at the absorption dip marked by P$_2$, respectively.}
\end{figure}

Fig. \ref{fig4}(a) describes the absorption spectrum of the hybrid structure of TiO$_2$ metasurface with monolayer WS$_2$ on top. Two peaks are located at 612.14 nm and 626.97 nm, respectively. The dip located at 620.06 nm shows that original resonance wavelength (616.2 nm) disappears, and the small red shifts of resonance location due to the large real part of the permittivity of WS$_2$ monolayer can be shown in Fig. \ref{fig2}(a) blue line. The obvious spectral splitting with two peaks and one dip as a result of the strong coupling between bare monolayer WS$_2$ and quasi-BIC, which indicates that coherent energy exchange is conducted between excitons and quasi-BIC. This finding can be explained by two-level coupled oscillator model\cite{Li2019,Li2019a,Qing2018}, as depicted in Fig. \ref{fig4}(b). The incident light can be regarded as ground state with the energ $E_0$. When the metasurfaces have proper geometrical parameters, a magnetic dipole with the energy 
$E_{\rm{MD}}$ can be excited by the incident light. The process can be thought of photon transition from the ground state to one excited state. In the same way, the interaction between photon and exciton in the monolayer WS$_2$ is considered as the energy transition process from $E_0$ to $E$exc. Furthermore, coherent energy exchange will occur between the magnetic dipole and exciton as they share the same energy. When the energy exchange rate is greater than each decay rate, the strong coupling happens, and the original two independent energy levels will be hybridized to form a new hybrid state named polariton with two new energy levels. The electric field distributions of the new hybrid state are shown in Figs. \ref{fig4}(c), \ref{fig4}(d) and  \ref{fig4}(e). Comparing the electric field distributions at the absorption peaks and dip, it is found that the local electric field at peaks are much higher than that at the dip, which further proves that original energy state around 2.014 eV (616 nm) disappears and forms two new state are 612.14 nm and 626.97 nm, respectively. Thus, the evident spectral splitting make clear that strong coupling between MD and exciton can be obtained by our design.
\begin{figure}[htbp]
	\centering
	\includegraphics
	[scale=0.32]{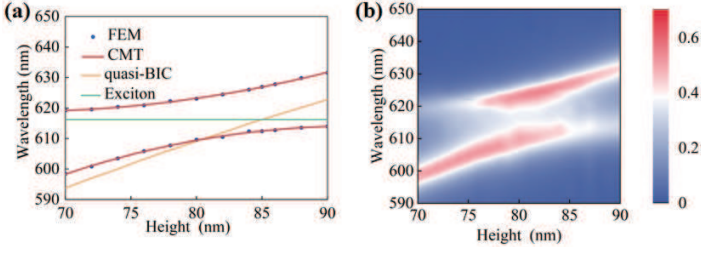}
	\caption{\label{fig5}\ (a) The wavelength of two new hybrid sates as a function of thickness ($L_2$=280nm). The green curve and dashed dots correspond to theoretical and numerical simulation, respectively, and the black line and red line depict the individual WS$_2$ monolayer and quasi-BIC modes, respectively. (b) The absorption spectra of the new hybrid state with different thickness.}
\end{figure}

It can be seen from the Fig. \ref{fig5}(a) (red line) that the resonant position of WS$_2$ does not change with thickness $H$, while the resonant wavelength of quasi-BIC shows a linear growth relationship with thickness $H$. Moreover, as the resonant wavelength increases, it will shift across the exciton resonant wavelength, therefore two branches of anti-crossover behavior can be captured and named lower branch (LB) and upper branch (UB), which is depicted from Fig. \ref{fig5}(b) the absorption spectrum of the hybrid structure with different thickness $H$. This can be explained by using coupled-mode theory (CMT). For in-plane vectors the eigenstates can be described as\cite{Liu2014,Deng2010}
\begin{small} 
	\begin{equation}
		\left[ \begin{array}{l}\label{eq2}
			{E_{\rm{q - BIC}}} + i\gamma _{\rm{q -BIC}}\\
			g
		\end{array} \right.\left. \begin{array}{l}
			g\\
			E_{exc} + i\gamma _{\rm{exc}}
		\end{array} \right]\left( \begin{array}{l}
			\alpha \\
			\beta 
		\end{array} \right) =  E_{LB,UB}\left( \begin{array}{l}
			\alpha \\
			\beta 
		\end{array} \right),
	\end{equation}
\end{small}
where, ${E_{{\rm{q - }}BIC}}$ and ${\gamma _{{\rm{q - }}BIC}}$ represent the quasi-BIC energy and dissipation, respectively. ${E_{{\rm{exc}}}}$ and ${\gamma _{{\rm{exc}}}}$  represent the energy and nonradiative decay rate of the uncoupled exciton, respectively, and $g$ is the coupling strength. $\alpha $ and $\beta $ are the Hopfield coefficients that describe the weighting of the quasi-BIC and Exciton for LB and UB, which should satisfy 
						${\left| \alpha  \right|^2} + {\left| \beta  \right|^2} = 1$. $E_{LB,UB}$ represent the eigvenvalues, which can be obtained from Eq. (\ref{eq2})
\begin{equation}
\begin{array}{l}
	{E_{LB,UB}} = \frac{1}{2}\left[ {{E_{{\rm{exc}}}}{\rm{ + }}{E_{{\rm{q}} - BIC}} + i({\gamma _{{\rm{exc}}}} + {\gamma _{{\rm{q}} - BIC}})} \right]\\
	\\
	\pm \sqrt {{g^2} + \frac{1}{4}\left[ {{E_{{\rm{exc}}}}{\rm{ - }}{E_{{\rm{q}} - BIC}} + i{{({\gamma _{{\rm{exc}}}} - {\gamma _{{\rm{q}} - BIC}})}^2}} \right]} 
\end{array}.\label{eq3}
\end{equation}

When the detuning $\Delta {\rm{ = }}{E_{{\rm{q - }}BIC}}{\rm{ - }}{E_{{\rm{exc}}}}{\rm{ = }}0$,  Eq. (\ref{eq3}) become
\begin{equation}
\begin{array}{l}
	{E_{LB,UB}} = \frac{1}{2}\left[ {{E_{{\rm{exc}}}}{\rm{ + }}{E_{{\rm{q}} - BIC}} + i({\gamma _{{\rm{exc}}}} + {\gamma _{{\rm{q}} - BIC}})} \right]\\
	\\
	\pm \sqrt {{g^2}{\rm{ - }}\frac{1}{4}\left[ {{{({\gamma _{{\rm{exc}}}} - {\gamma _{{\rm{q}} - BIC}})}^2}} \right]} 
\end{array}.\label{eq4}
\end{equation}

Then we obtain the Rabi splitting energy 
\begin{equation}
\hbar \Omega  = 2\sqrt {{{\rm{g}}^2}{\rm{ - }}{{({\gamma _{{\rm{q - BIC}}}}{\rm{ - }}{\gamma _{{\rm{exc}}}})}^2}/4},\label{eq5}
\end{equation}
which is owning to the strong coupling between quasi-BIC and exciton. Here, we also calculate the ${\gamma _{{\rm{q - }}BIC}}$  =14.93 meV and 
${\gamma _{{\rm{exc}}}}$  =15 meV from  Fig. \ref{fig3}(a), the Rabi splitting energy
$\hbar \Omega $=46 meV can be extracted from FEM simulation results shown in Fig. \ref{fig5}(a) (dashed line), satisfying the condition of strong coupling (
$\hbar \Omega  > {{\left( {{\gamma _{{\rm{q - }}BIC}} + {\gamma _{{\rm{exc}}}}} \right)} \mathord{\left/
		{\vphantom {{\left( {{\gamma _{{\rm{q - }}BIC}} + {\gamma _{{\rm{exc}}}}} \right)} 2}} \right.
		\kern-\nulldelimiterspace} 2}$ ). We then compare the dissipation rate with the coupling strength $g$. From Eq. (\ref{eq5}), we obtained g=23 meV, which indicated
$g > {{\left| {{\gamma _{{\rm{exc}}}} - {\gamma _{{\rm{q - }}BIC}}} \right|} \mathord{\left/
{\vphantom {{\left| {{\gamma _{{\rm{exc}}}} - {\gamma _{{\rm{q - }}BIC}}} \right|} 2}} \right.
\kern-\nulldelimiterspace} 2}$  and 
$g > \sqrt {{{\left( {{\gamma _{{\rm{exc}}}}^2 + {\gamma _{{\rm{q - }}BIC}}^2} \right)} \mathord{\left/
	{\vphantom {{\left( {{\gamma _{{\rm{exc}}}}^2 + {\gamma _{{\rm{q - }}BIC}}^2} \right)} 2}} \right.
	\kern-\nulldelimiterspace} 2}}$ . These results are a further proof that we are indeed in the strong coupling regime.
\begin{figure}[htbp]
\centering
\includegraphics
[scale=0.45]{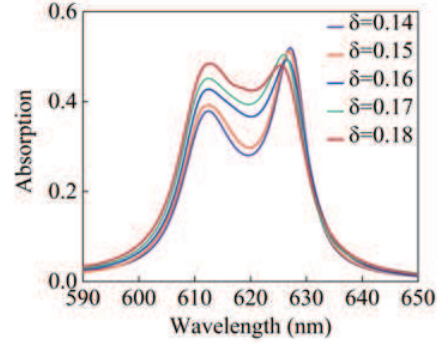}
\caption{\label{fig6}\ The absorption curves of two new hybrid sates with a variable short nanobar, resonant wavelength all at 616.2 nm by tuning the thickness.}
\end{figure}

Fig. \ref{fig6} shows the absorption spectra of quasi-BIC and exciton coupling with the different asymmetric parameters but at the same resonant wavelength by tuning the thickness $H$,                                                                                                                                     which indicates that coupling strength g will reduce with the decrease of the asymmetric parameter. It can be described by CMT that when the dissipation loss of the quasi-BIC resonance gets close to the nonradiative decay rate of the exciton, the Rabi splitting reaches its maximum\cite{Deng2010,Piper2014,Xiao2020}. For the smaller asymmetric parameters, the larger local electric field, accompanied by narrower line width and the dissipation loss of quasi-BIC mode, as a result of limiting the total number of photons related to the interaction with excitons. Therefore, it is important to find a balance between local electric field and spectral line width.
\begin{figure}[htbp]
	\centering
	\includegraphics
	[scale=0.35]{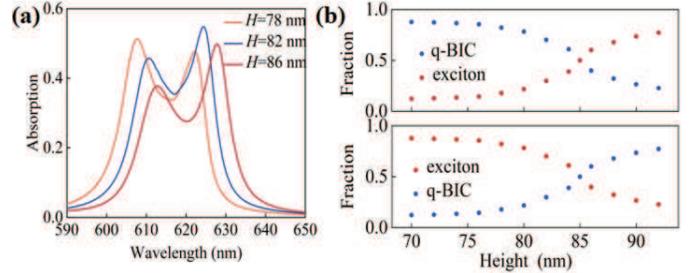}
	\caption{\label{fig7}(a) The absorption curves of two new hybrid sates with a variable thickness $H$, but a fixed length $L_2$=280 nm of short nanobar. (b) The fraction curve of exciton (red dots) and quasi-BIC (black dots) in the UB and LB, respectively.}
\end{figure}

Finally, we also study the absorption curves of two new hybrid sates with a variable thickness $H$, but with fixed asymmetric parameters shown in Fig. \ref{fig7}(a). It is found that the absorption peaks of LB increases while the absorption peaks of UB decreases with the decrease of thickness $H$, which can be explained by the relative weightings of exciton and quasi-BIC in new hybrid state.

 The weighting of the quasi-BIC and exciton constituents in the LB and UB can be drived from Eq. (\ref{eq2}) and the fractions for the LB/UB polariton are 
\begin{eqnarray}
 	{\left| \alpha  \right|^2}{\rm{ = }}\frac{1}{2}\left( {1 \pm \frac{\Delta }{{\sqrt {{\Delta ^2}{\rm{ + }}4{g^2}} }}} \right),\label{eq6}\\ 	
{\left| \beta  \right|^2}{\rm{ = }}\frac{1}{2}\left( {1 \mp \frac{\Delta }{{\sqrt {{\Delta ^2}{\rm{ + }}4{g^2}} }}} \right),\label{eq7}
\end{eqnarray}
 which are shown in Fig. \ref{fig7}(b). We can found that as the thickness $H$ increase, the exciton (quasi-BIC) fraction increases in UB (LB) and decreases in LB (UB). In other words, with the decrease of thickness in the LB, the weight of excitons decreases, which means that the number of excitons participate in the coupling decreases and the absorption summit of the lower branch decreases.
 \section{\label{sec3}Conclusion}
 In conclusions, we have theoretically investigated the strong coupling between the WS$_2$ excitons and quasi- BIC mode supported by TiO$_2$ metasurfaces. The Rabi splitting energy up to 46.86 meV is observed in the absorption spectrum of the hybrid structure. Furthermore, anticrossing behavior as a typical feature of strong coupling can be achieved by tuning the asymmetric parameters and the thickness of TiO$_2$ metasurface. More importantly, it is found that the line width of the quasi-BIC mode and local electric field enhancement should be balanced since both of them affect the strong coupling. Beyond this work, the proposed configuration can be extended to diverse kinds of strong coupling system, in principle, with various dielectric metasurface designs and different TMDCs (MoS$_2$, WSe$_2$ etc.). Therefore, this paper provides a strategically important method for metasurface-enhanced strong coupling, and offers designable, low-energy consumption, practical platform for future research of quantum phenomena and nanophotonic devices.
\begin{acknowledgments}	
This work is supported by the National Science Foundation of China (Grants No. 12064025, 11947065), Natural Science Foundation of Jiangxi Province (Grant No. 20202BAB211007), Jiangxi Provincial Cultivation Program for Academic and Technical Leaders of Major Subjects and the Interdisciplinary Innovation Fund of Nanchang University (Grant No. 2019-9166-27060003).

\end{acknowledgments}

%

\end{document}